\documentclass[twocolumn,pra,showpacs]{revtex4}

\begin{document}


\title{Limits on a Dynamically Varying Fine Structure Constant }

\author{E.J. Angstmann}
\affiliation{School of Physics, University of New South Wales,
Sydney 2052, Australia}
\author{V.V. Flambaum}
\affiliation{School of Physics, University of New South Wales,
Sydney 2052, Australia}

\author{S.G. Karshenboim} 
\affiliation{D. I. Mendeleev Institute for Metrology (VNIIM), St. Petersburg 198005, Russia\\
Max-Planck-Institut f\"ur Quantenoptik, 85748 Garching, Germany
}

\date{\today}

\begin{abstract}
We show that using the modified form of the Dirac Hamiltonian as suggested by Bekenstein does not affect the analysis in \cite{Webb,Srianand,Quast,Dzuba99} of QSO data pertaining to a measurement of $\alpha$ variation. We obtain the present time limit on Bekenstein's parameter, $\tan^{2}\chi =(0.2 \pm 0.7)\times10^{-6}$, from the measurement of the hydrogen $2p$ fine structure using value of $\alpha$ obtained from different experiments.

\end{abstract}

\maketitle

\section{Introduction}
The possible variation of the fine structure constant, $\alpha$, 
is currently a very popular research topic. Webb {\textit{et al.}}
 \cite{Webb} have found evidence of $\alpha$ 
variation by analyzing absorption lines in QSO spectra, while other 
groups \cite{Srianand,Quast} have used the same method \cite{Dzuba99} 
but found no evidence of $\alpha$ variation. Recently Bekenstein 
\cite{Bekenstein} has questioned the validity 
of the analysis used by these groups. Bekenstein shows that within the 
framework of dynamical $\alpha$ variability the form of the Dirac 
Hamiltonian relevant for an electron in an atom departs from the 
standard form. Unfortunately no self consistent quantum 
electrodynamic theory was derived. Instead the Dirac 
Hamiltonian, \^H, was presented for an electron bound by a 
Coulomb field:  
\begin{eqnarray}\label{eq:H}
\hat{H} & = & \hat{H}_{0}+\delta\hat{H} \\
\label{eq:H_0}\hat{H}_{0} & = & (-\imath\hbar c\mbox{\boldmath$\alpha$}\cdot\mbox{\boldmath$\nabla$}+mc^2\beta+e\Phi I)  \\
\label{eq:deltaH}\delta\hat{H} & = & (I-\beta)\tan^2\chi\cdot V_{C}
\end{eqnarray}
where $V_{C}=-Ze^{2}/r$. The last term is related to an
 effective correction to the Coulomb field due to the dynamical 
nature of $\alpha$, and $\tan^{2}\chi$ is a small parameter. In 
other words, a dynamically varying fine structure constant can be 
accounted for as a perturbation of the Dirac Hamiltonian. This 
perturbative term, $\delta\textrm{\^H}$, vanishes in a non-relativistic 
approximation but produces some relativistic corrections which 
can be studied both in astrophysical spectra and laboratory conditions.

In the following sections we will show how this perturbation shifts 
atomic energy levels. In particular we pay attention to heavy atoms 
which provide us with astrophysical data and are the most sensitive to a 
possible $\alpha$ variation. We also consider atomic hydrogen, 
since it is the best understood atomic system for laboratory experiments.

\section{Multielectron Atoms}
Multielectron atoms are of interest to us since they are the most sensitive to a 
varying $\alpha$. We performed a calculation to show how the modified
 form of the Dirac Hamiltonian affects the energy of an external electron 
in a heavy atom. We averaged $\delta$\^H, presented in Eq. 
(\ref{eq:deltaH}), over the relativistic wave 
function for electrons near the nucleus \cite{Khriplovich} at 
zero energy. The use of the wave function for the electrons 
at zero energy is justified by noting that the main contribution 
to the matrix element of $\delta \hat{H}$ comes from distances close to the 
nucleus, $r \sim a/Z$. At a distance of one Bohr radius, 
$r \sim a$, the potential is screened by the other electrons 
and the potential energy is given by $V_{C} \sim -mc^{2}\alpha^{2}$. 
This is of the same order of magnitude as the binding energy 
of the electron, $E \sim -mc^{2}\alpha^{2}$. However, close to the 
nucleus, at $r \sim a/Z$, screening is negligible and $
V_{C}\sim -Z^{2}mc^{2}\alpha^{2}$. Inside this region $|V_{C}|\gg |E|$ 
and so the binding energy of the electron can be safely ignored. 
A detailed derivation of the $Z^{2}$ enhancement of $<\delta \hat{H}>$
 for $r \lesssim a/Z$ is provided in the appendix.

The non-relativistic limit of $\delta E$ was taken. This gave:
\begin{equation}\label{eq:heavyatomcorrect}
\frac{\delta E}{E} = \frac{2(Z\alpha)^{2}\tan^{2}\chi}{\nu}\frac{1}{j + 1/2}
\end{equation}
where $E=-Z_{a}^{2}mc^{2}\alpha^{2}/(2\nu^{2})$ is the energy of the electron, and $Z_{a}$ is the charge ``seen'' by the electron - it is 1 for atoms, 2 for singly charged ions etc. The derivation of Eq. (\ref{eq:heavyatomcorrect}) 
can be found in the appendix.

It is interesting to note that this correction to the energy of
 the electron has exactly the same form as the relativistic 
correction, $\Delta$, to the energy of an external electron:
\begin{equation}\label{eq:relcorrect}
\frac{\Delta}{E} = \frac{(Z\alpha)^{2}}{\nu}\frac{1}{j + 1/2}.
\end{equation}
We can sum up Bekenstein's relativistic correction, Eq. (\ref{eq:heavyatomcorrect}), and the usual relativistic correction to give
\begin{equation}
\frac{\Delta'}{E} \simeq \frac{(Z\alpha')^{2}}{\nu}\frac{1}{j + 1/2}.
\end{equation}
where $\alpha'=\alpha(1+\tan^{2}\chi)$.

The works \cite{Webb,Srianand,Quast} used a method suggested in \cite{Dzuba99} for the analysis of absorption lines. A comparison between different frequencies is used. In this method only the relativistic corrections, $\Delta'/E$, are used to determine $\alpha$ variation since any variation in the energy in the non-relativistic limit is absorbed into the redshift parameter and also scales the same way for all elements. Since Eq. (\ref{eq:heavyatomcorrect}) and (\ref{eq:relcorrect}) 
are directly proportional the effect of the modified form of the Dirac 
Hamiltonian is indistinguishable from a small change in $\alpha^{2}$ in Eq. (\ref{eq:relcorrect}). A measurable change in $\alpha$ would in fact be a change of $\alpha'=\alpha(1+\tan^{2}\chi)$. The astrophysical data can not distinguish between $\alpha$ and $\tan^{2}\chi$ variation.

 Note that the proportionality of $\delta E$  and  $\Delta$ has a simple
 explanation. The relativistic corrections to the Schr\"odinger Hamiltonian
(e.g. the spin-orbit interaction) and the non-relativistic limit
of $\delta\hat{H}$ have similar dependence on $r$ and both are proportional
to the electron density $\psi ^2$ at $r \sim a/Z$. 
The proportionality of $\delta E$  and  $\Delta$
also holds for high orbitals (small binding energy) in the
 pure Coulomb case (see next section).
 
In this derivation we assumed 
that we could consider the unscreened Coulomb field, this is 
clearly not the case for a valence electron in a many electron 
atom. We justify this assumption by once again noting that the main 
contribution to $\delta E$ is given by distances close to the 
nucleus, $r \sim a/Z$.
 At this distance the 
main screening comes from the $1s$ electrons and we can use 
Slater's rules to estimate the screening corrections to Eq. 
(\ref{eq:heavyatomcorrect}) and (\ref{eq:relcorrect}):
\begin{eqnarray}
\frac{\delta E}{E} & = &  \frac{2\alpha^{2}(Z-0.6)^{2}\tan^{2}\chi}{\nu}\frac{1}{j + 1/2} \\
\frac{\Delta}{E} & = & \frac{\alpha^{2}(Z-0.6)^{2}}{\nu}\frac{1}{j + 1/2}.
\end{eqnarray}
 This does not 
affect the proportionality of the two terms and makes 
very little difference to the results in heavy atoms ($\sim 1/Z$).
The correction from the non-zero energy of the external electron
is even smaller ($\sim 1/Z^2$).
Consideration of many-body correlation corrections has shown that 
this does not change the proportionality relationship either. 
The point is that the expressions for the correlation corrections
obtained using the many-body perturbation theory (or the configuration
interaction method) contain the single-particle matrix elements
of  $\delta\hat{H}$ and that of the relativistic corrections which
are proportional to each other. This makes the final results
proportional. Because of this proportionality it is not possible to 
derive values for $\alpha$ and $\tan^{2}\chi$ separately in 
mulitelectron atoms. However, a separation of these values can be 
achieved in hydrogen.

\section{Calculations Involving Hydrogen Atom}

The case is somewhat simplified for the hydrogen atom and other
 hydrogen-like ions as there are no inter-electron interactions. 
There are also very accurate experimental measurements of 
transition frequencies in hydrogen.

We confirm Bekenstein's result \cite{Bekenstein} that applying 
the Hamiltonian (\ref{eq:H}) one can derive:
\begin{equation} \label{eq:deltaE}
\delta E = -\frac{mc^{2}Z^{4}\alpha^{4}}{n^{3}}\left(\frac{1}{j+1/2}-\frac{1}{2n}\right) \tan^{2}\chi \;.
\end{equation}
The relativistic correction to the electron energy is
\begin{equation}\label{eq:delta}
\Delta = -\frac{mc^{2}Z^{4}\alpha^{4}}{2n^{3}}\left(\frac{1}{j+1/2}-\frac{3}{4n}\right).
\end{equation}
Note that for large $n$ (zero energy), $\delta E$ is again proportional 
to the relativistic correction, $\Delta$.

It is possible to obtain a limit on the $\tan^{2}\chi$ parameter by comparing the theoretical and experimental data for the $2p_{3/2}-2p_{1/2}$ splitting. The experimental value
\begin{equation}\label{eq:experimental}
f_{2p_{3/2}\rightarrow2p_{1/2}}({\rm exp})=10\,969\,045(15)\, \textrm{kHz}
\end{equation}
is derived from two experimental results,
\begin{eqnarray}
f_{2s_{1/2}\rightarrow2p_{3/2}}({\rm exp}) & = & 9\,911\,200(12)\,\textrm{kHz}\\
f_{2p_{1/2}\rightarrow2s_{1/2}}({\rm exp}) & = & 1\,057\,845(9)\, \textrm{kHz}
\end{eqnarray}
presented in \cite{Lundeen} and \cite{Hagley} respectively. It has to be compared with a theoretical value which we take from a compilation \cite{Mohr} (see also review \cite{Eides})
\begin{equation}\label{eq:theoretical}
f_{2p_{3/2}\rightarrow2p_{1/2}}({\rm theory})=10\,969\,041.2(15)\, \textrm{kHz}\;.
\end{equation}
Noting  that $\delta E$ has not been accounted for in the Eq. (\ref{eq:theoretical}), but will be present in Eq. (\ref{eq:experimental}), and using Eq. (\ref{eq:deltaE}) we can write:
\begin{equation}
\frac{mc^{2}\alpha^4}{16h}\tan^{2}\chi = f_{2p_{3/2}\rightarrow2p_{1/2}}({\rm exp})-f_{2p_{3/2}\rightarrow2p_{1/2}}({\rm theory}).
\end{equation}
The leading contribution to $f_{2p_{3/2}\rightarrow2p_{1/2}}(\rm theory)$ is given by $mc^{2}\alpha^4/32h$, this allows us to write:
\begin{equation}
f_{2p_{3/2}\rightarrow2p_{1/2}}({\rm exp})=f_{2p_{3/2}\rightarrow2p_{1/2}}({\rm theory})(1+2\tan^{2}\chi)
\end{equation}
Using the values above we can obtain the limit $\tan^{2}\chi=2(7)\times10^{-7}$.
It is assumed here that $f_{2p_{3/2}\rightarrow2p_{1/2}}({\rm theory})$ is expressed
in terms of $\alpha$ which is extracted from the measurements of  parameters
which are not sensitive to $\tan^{2}\chi$  (i.e. they depend on $\alpha$ rather than
 on $\alpha'$). Indeed, one of the  values 
of $\alpha$  is derived via a complicated chain of 
relations with $\alpha$ eventually coming from the Rydberg constant 
which is quite weakly affected by $\delta\hat{H}$ (the relative
value of the correction is  of order of $\alpha^2\tan^{2}\chi$
since the matrix element of $\delta\hat{H}$ vanishes in 
a leading non-relativistic approximation). The 
most accurate result obtained this way is $\alpha^{-1}=137.036\,
000\,3(10)$ \cite{chu}.

A self consistent quantum electrodynamic theory with a dynamically 
varying $\alpha$ should meet some even stronger constraints 
due to a comparison of the value of the fine structure constant 
from the anomalous magnetic moment of the electron ($\alpha^{-1}=
137.035\,998\,80(52)$ \cite{kinoshita}) with the Rydberg constant
value.  Such a comparison will likely lead to a limitation on 
$\tan^{2}\chi$ at a level of a few parts in $10^{-8}$ since 
$\delta \alpha/\alpha = 11(8)\times 10^{-9}$, from comparison of 
the values for $\alpha^{-1}$ given above.

Before any modification of QED due to a varying $\alpha$ is considered seriously another set of questions need to be answered. These questions should target its gauge invariance, renormalizability and Ward identities, which supports the same charge for electrons and protons. The current QED construction is quite fragile and it is not absolutely clear if it can be successfully extended.

\section{Conclusions}

In conclusion, using the modified form of the Dirac Hamiltonian Eqs.
(\ref{eq:H})-(\ref{eq:deltaH}) does not 
affect the analysis used in \cite{Webb,Srianand,Quast}. They
 measure the variation of
 $\alpha'=\alpha(1+\tan^{2}\chi)$.
The present time limit  $\tan^{2}\chi=(0.2 \pm 0.7)\times10^{-6}$ is obtained from the measurement
of the hydrogen $2p$ fine structure using value of $\alpha$ obtained from different experiments.
Note that according to \cite{Webb} the value of $\alpha'$ was smaller in the past, the
 last measurement gave
  $\Delta\alpha'/\alpha' = (-0.54 \pm 0.12)\times10^{-5}$. If there is no other source
of variation of $\alpha$ this would require a negative value of $\tan^{2}\chi$ ($\tan^{2}\chi = (-0.52 \pm 0.14) \times 10^{-5}$)
since the present value of $\tan^{2}\chi$ is small. Actually, the choice of the integration constants in the Bekenstein paper precludes considering epochs with $\alpha'<\alpha$ \cite{Bekenstein}. However this should not be deemed a principle problem.

\section*{Acknowledgments}
The work of SGK was supported in part by RFBR grant 03-02-16843. A part of the work was done during a short but fruitful visit of VVF to Max-Plank Insitute for Quantum Optics (Garching). The work of EJA and VVF was supported by the Australian Research Council.

\appendix
\section*{Appendix - Calculation of $\delta E$}
The most convenient way to calculate $\delta E$ is to calculate the matrix element of the operator $\delta\hat{H} = (I-\beta)\tan^2\chi\cdot V_{C}$ for an external electron in a many-electron atom or ion using a relativistic wave function. In order to use the relativistic wave functions for electrons near the nucleus at zero energy it is necessary to demonstrate that the major contribution to the matrix element of $\delta \hat{H}$ comes from distances $r \lesssim a/Z$, where the screening of the nuclear potential and the external electron energy can be neglected. Only the contribution at distances $r \lesssim a/Z$ have a $Z^{2}$ enhancement, the contribution at $r \sim a$ does not have this enhancement since the atomic potential at this distance is screened, $V_{C} \sim e^{2}/r$, and has no $Z$ dependence.

To demonstrate the $Z^{2}$ enhancement of the $r\lesssim a/Z$ contribution let us consider the non-relativistic limit of the operator $\delta \hat{H}$. The matrix 
\begin{displaymath}
I-\beta =  \left( \begin{array}{cc} 0 & 0 \\ 0 & -2 \end{array} \right)
\end{displaymath}
has only lower components, it follows that the matrix element,
\begin{equation}
\psi^{+}(I-\beta)V_{C}\psi = -2\chi^{+}V_{C}\chi,
\end{equation}
where 
\begin{displaymath}
\psi = \left( \begin{array}{c} \varphi \\ \chi \end{array}\right)
\end{displaymath}
is the Dirac spinor. In the non-relativistic limit 
\begin{displaymath}
\chi = \frac{\mbox{\boldmath $\sigma$}\cdot\mathbf{p}}{2mc}\varphi
\end{displaymath}
and this gives
\begin{eqnarray}\label{eq:deltaHmatrix}
\delta \hat{H} & = & -\frac{1}{2m^{2}c^{2}}(\mbox{\boldmath $\sigma$}\cdot\mathbf{p})V_{C}(\mbox{\boldmath $\sigma$}\cdot\mathbf{p}) \tan^{2}\chi\nonumber\\
 & \approx &\Big( -V_{c}\frac{p^{2}}{2m^{2}c^{2}}+\frac{i\hbar}{2m^{2}c^{2}}(\nabla V_{C}\cdot\mathbf{p})- {}\nonumber\\ & & {} -\frac{\hbar^{2}}{2m^{2}c^{2}}\frac{1}{r}\frac{dV_{C}}{dr}\mbox{\boldmath $\sigma$}\cdot\mathbf{l} \Big) \tan^{2}\chi.
\end{eqnarray}
This derivation is similar to the standard derivation of the spin-orbit interaction term in the non-relativistic expansion of the Dirac Hamiltonian. Let us now compare the contributions of $r\lesssim a/Z$ and $r\sim a$ to the matrix element of $\delta\hat{H}$. Consider, for example, the last spin-orbit term in Eq. (\ref{eq:deltaHmatrix}) which is proportional to the usual spin-orbit interaction. The electron wave function at $r \sim a/Z$ is given by $\varphi^{2} \sim Z/a^{3}$ \cite{Landau}, the spin-orbit operator is proportional to
\begin{displaymath}
\frac{1}{r}\frac{dV_{C}}{dr}\sim\frac{1}{r^{3}}Ze^{2}
\end{displaymath}
and the integration volume is proportional to $ r^{3}$. As a result we can write:
\begin{equation}
<\delta\hat{H}>\sim Z^{2}\frac{\hbar^{2}e^{2}}{2m^{2}c^{2}a^{3}}.
\end{equation}
For $r\sim a$, the wave function is $\varphi^{2} \sim 1/a^{3}$, the spin-orbit operator is proportional to 
\begin{displaymath}
\frac{1}{r}\frac{dV_{C}}{dr}\sim\frac{1}{r^{3}}e^{2}
\end{displaymath}
and the integration volume is still proportional to $r^{3}$. Therefore 
\begin{equation}
<\delta\hat{H}>\sim \frac{\hbar^{2}e^{2}}{2m^{2}c^{2}a^{3}},
\end{equation}
this is $Z^{2}$ times smaller than at $r \lesssim a/Z$. The same conclusion is also valid for the first two terms in Eq. (\ref{eq:deltaHmatrix}). This estimate demonstrates that the main contribution to $\delta \hat{H}$ comes from small distances $r\lesssim a/Z$. This conclusion is similar to that for the relativistic corrections to atomic electron energy.

We perform the actual calculation of the matrix element of $\delta\hat{H}$ using the relativistic Coulomb wave functions for zero-energy electrons near the nucleus. These can be expressed in terms of Bessel functions as \cite{Khriplovich}:
\begin{eqnarray}\label{bessel}
f_{njl}(r) & = & \frac{c_{njl}}{r}\Big((\gamma+\kappa)J_{2\gamma}(x)-\frac{x}{2}J_{2\gamma-1}(x)\Big) \nonumber\\
g_{njl}(r) & = & \frac{c_{njl}}{r}Z\alpha J_{2\gamma}(x)
\end{eqnarray}
where $x = (8Zr/a)^{1/2}$, $\gamma = \sqrt{(j+1/2)^{2}-Z^{2}\alpha^{2}}$, $\kappa = (-1)^{j+1/2-l}(j+1/2)$ and 
\begin{displaymath}
c_{njl}=\frac{\kappa}{|\kappa|}\Big(\frac{1}{Za\nu^{3}}\Big)^{1/2}Z_{a}.
\end{displaymath}
$\delta E$ can now be calculated using these wave functions and $\delta \hat{H}$:
\begin{eqnarray*}
\delta E & = & \int \psi^{+}\delta \hat{H} \psi dV \nonumber\\
 & = & -2\tan^2{\chi}\int_{0}^{\infty}g_{njl}^{+}V_{c}g_{njl}r^{2}dr \nonumber\\
 & = & -2e^{2}Z^{3}\alpha^{2}\tan^{2}\chi c_{njl}^{2}\int_{0}^{\infty}J_{2\gamma}^{2}(x)\frac{dr}{r} \\
 & = & -4e^{2}Z^{3}\alpha^{2}\tan^{2}\chi c_{njl}^{2}\int_{0}^{\infty}J_{2\gamma}^{2}(x)\frac{dx}{x}.
\end{eqnarray*}
We now use the relationships between Bessel functions and Gamma functions to write this as:
\begin{eqnarray*}
\delta E &  = & -2e^{2}Z^{3}\alpha^{2}\tan^{2}\chi c_{njl}^{2}\frac{\Gamma(2\gamma)}{\Gamma(2\gamma+1)} \\
 & = & -\frac{e^{2}Z^{3}\alpha^{2}\tan^{2}\chi c_{njl}^{2}}{\gamma}.
\end{eqnarray*}
When we substitute in for $c_{njl}$ we obtain:
\begin{displaymath}
\delta E = -\frac{mc^{2}Z^{2}Z_{a}^{2}\alpha^{4}\tan^{2}\chi}{\nu^{3}\gamma}.
\end{displaymath}
Finally, we take the non-relativistic limit by replacing $\gamma$ with $j+1/2$,
\begin{equation}\label{eq:derivedE}
\delta E = - \frac{mc^{2}Z^{2}Z_{a}^{2}\alpha^{4}\tan^{2}\chi}{\nu^{3}(j+1/2)}.
\end{equation}
To obtain Eq. (\ref{eq:heavyatomcorrect}) we simply divide Eq. (\ref{eq:derivedE}) by $E=-Z_{a}^{2}m\alpha^{2}/(2\nu^{2})$. Note that the spin-orbit contribution to Eq. (\ref{eq:relcorrect}) can be obtained in an analogous manner.

   Finally, we should present a very simple derivation of Eqs. 
(\ref{eq:heavyatomcorrect}) and (\ref{eq:relcorrect}) based on the
 results obtained for the pure Coulomb case, see Eqs. (\ref{eq:deltaE})
and (\ref{eq:delta}). For the high electron orbitals ($n>>1$)
the electron energy  at $r \sim a/Z$ may be neglected in comparison with
the Coulomb potential and   
the Coulomb results Eqs. (\ref{eq:deltaE})
and (\ref{eq:delta}) are proportional to the electron density at
$r \sim a/Z$  where $\psi^2 \sim 1/n^3$.
 For the external electron in heavy atoms
the situation is similar. The external electron wave function in  Eqs.
 (\ref{bessel}) at  $r \sim a/Z$ is proportional to the
 Coulomb wave function for small energy ($n>>1$).
Therefore, to find the matrix elements for the external electron
 we should take the Coulomb results  Eqs. (\ref{eq:deltaE})
and (\ref{eq:delta}) and multiply them by the ratio of the electron densities
for the external electron in the many-electron atom and the Coulomb electron.
This immediately gives  Eqs. 
(\ref{eq:heavyatomcorrect}) and (\ref{eq:relcorrect}).

\end{document}